\RequirePackage{fix-cm}
\documentclass[smallextended]{svjour3}       
\smartqed  
\usepackage{graphicx}
\usepackage{amsmath}
\usepackage{url}
\usepackage{subfigure}
\usepackage{comment}
\usepackage[numbers]{natbib}
\usepackage{hyperref}
\newcommand{\br}[1]{{\left(#1\right)}}
\newcommand{\vct}[1]{\underline{#1}}
\newcommand{\prt}[3][\empty]{\partial_{#2}^{#1}#3}
\newcommand{\eq}{^{\mathrm{eq}}}
\def\pos{\vct{x}}
\def\vel{\vct{u}}
\def\ldir{\vct{c}_i}
\def\fpdf{f_i}
\def\tpdf{h_i}
\providecommand{\phone}[1]{Tel.: #1}
\providecommand{\fax}[1]{Fax.: #1}
\newcommand{\walberla}{\textsc{waLBerla}}
\DeclareMathAlphabet{\mathpzc}{OT1}{pzc}{m}{it}
\newcommand{\pe}{$\mathpzc{pe}$}
\journalname{The International Journal of Advanced Manufacturing Technology}
\begin{document}
\title{Numerical Investigations on Hatching Process Strategies for Powder Bed Based Additive Manufacturing using an Electron Beam}
\titlerunning{Numerical Investigations on Hatching Process Strategies}
\author{Matthias Markl$^*$ \and Regina Ammer \and Ulrich R\"{u}de \and Carolin K\"{o}rner}
\institute{M. Markl$^*$, C. K\"orner \at Chair of Metals Science and Technology, Friedrich-Alexander-Universit\"at Erlangen-N\"urnberg, Martensstr. 5, 91058 Erlangen, Germany \\ 
           $^*$ \email{matthias.markl@fau.de}, \phone{+49-9131-8520783}, \fax{+49-9131-8527515}
      \and R. Ammer, U. R\"ude \at Chair for System Simulation, Friedrich-Alexander-Universit\"at Erlangen-N\"urnberg, Cauerstr. 11, 91058 Erlangen, Germany}
\date{Received: March 17, 2014 / Accepted: date}
%
\maketitle
\begin{abstract}
This paper investigates in hatching process strategies for additive manufacturing using an electron beam by numerical simulations.
The underlying physical model and the corresponding three dimensional thermal free surface lattice Boltzmann method of the simulation software are briefly presented.
The simulation software has already been validated on the basis of experiments up to 1.2\,kW beam power by hatching a cuboid with a basic process strategy,
whereby the results are classified into `porous', `good' and `uneven', depending on their relative density and top surface smoothness.
In this paper we study the limitations of this basic process strategy in terms of higher beam powers and scan velocities to exploit the future potential of high power electron beam guns up to 10\,kW.
Subsequently, we introduce modified process strategies, which circumvent these restrictions, to build the part as fast as possible under the restriction of a fully dense part with a smooth top surface.
These process strategies are suitable to reduce the build time and costs, maximize the beam power usage and therefore use the potential of high power electron beam guns.
\end{abstract}
%
%
\section{Introduction}
%
%
Selective electron beam melting (SEBM) is a powder bed based additive manufacturing (AM) technology used to produce metallic structures.
The great advantage of SEBM manufacturing is the construction of strong and flexible structures with a high geometric complexity, 
which opens new opportunities in many different applications, ranging from medical implants to aerospace parts.
\citeauthor{Heinl2007}~\cite{Heinl2007} describe the SEBM process and show its potential to manufacture cellular structures made of titanium alloys as medical implants.
This approach is extended by \citeauthor{Murr20111396}~\cite{Murr20111396} to fabricate patient specific knee implants with a high biocompatibility.
\citeauthor{Rawal2013}~\cite{Rawal2013} focus on the development of aerospace wave guide brackets and show the competitiveness and advantages of SEBM to conventional subtractive manufacturing.

In contrast to other AM technologies, \citeauthor{Vayre201289}~\cite{Vayre201289} conclude that the SEBM process produces parts with a sufficient surface smoothness, dimensional quality and material properties at relatively low build times.
However, until today the build rates are still not sufficient, that SEBM is competitive to conventional technologies in less specific, mass production applications.
In order to accelerate build rates and to ensure a reliable final quality, numerical simulations are used to improve state-of-the-art process strategies.

We discretize the SEBM process with a three dimensional thermal free surface lattice Boltzmann (LB) method, including physical phenomena, like hydrodynamic flow, capillarity, wetting, 
as well as beam absorption and phase transformations, but excluding viscous heat dissipation, compressibility and evaporation effects.
The implementation of the simulation software is embedded in the framework \walberla{}\footnote{\scriptsize\url{www.walberla.net}} (widely applicable lattice Boltzmann solver from Erlangen),
which is a lattice Boltzmann based fluid flow solver with highly parallelized kernels 
and the \pe{}\footnote{\scriptsize\url{www10.informatik.uni-erlangen.de/de/Research/Projects/pe/}} (physics engine) framework,
which performs the parallel simulation of the powder bed generation.
\citeauthor{walberla2011}~\cite{walberla2011} describe in detail the software concepts of \walberla{} including CPU-GPU coupling, software quality and performance aspects and a two-way coupling to the \pe{} framework.
\citeauthor{kostler2013cse}~\cite{kostler2013cse} explain the extended capabilities of the framework to simulate different applications in the field of computational science and engineering.

The numerical methods have already been validated to analytical solutions~\cite{Ammer2013}.
Additionally, a simulation software calibration focused on single line examples and a physical validation to hatching experiments with a basic process strategy has been performed~\cite{Ammer2014}. 
These results demonstrate the high accordance of the thermal free surface LB approach to the real SEBM process and allows us to analyze and improve different process strategies.

The main objective of this work is to develop process strategies, which are able to use high scan velocities and beam powers in order to reduce build times.
State-of-the-art SEBM machines provide up to 3\,kW beam power\footnote{Arcam Q10 and Q20 with 3\,kW beam power: \url{http://www.arcam.com/technology/products}}.
However, reliable experiments with the basic process strategy are only available up to 1.2\,kW \cite{Juechter2014252}.
Beyond this beam power, constant beam characteristics in terms of energy distribution and focus spot as well as exact trajectories are not provided.
\citeauthor{Kornilov2013843}~\cite{Kornilov2013843} analyze plasma electron guns with a higher beam power of 4\,kW and show their competitiveness to conventional cathodes. 
Extending the current research on plasma cathodes, future electron guns will provide a maximum beam power of at least 10\,kW.
Nowadays it is commonly known, that state-of-the-art process strategies cannot use this potential advantage, 
because high evaporation rates and resulting evaporation pressures will destroy the top surface smoothness of the part as well as its material composition.
In order to encourage the development of new process strategies and because of the current machine restrictions,
we have to restrict all investigations of this work on numerical simulations.

In a first step the process window of the basic process strategy is extended to determine the beam power and scan velocity limitations.
Subsequently, two different techniques are studied to improve the basic process strategy. 
On the one hand the beam diameter is increased and on the other hand the line offset is decreased.
With both strategies the total build time is decreased, by using higher scan velocities resulting in a high beam power usage.
This accelerates the process, reduces manufacturing costs and opens the opportunity for a larger variety of parts and applications.

The outline of the article is as follows: In the following section the physical model and the three dimensional thermal free surface LB method used for the simulation of the SEBM process are summarized.
Subsequently, the numerical simulation setup and classification criteria used for all numerical simulations are defined. 
In the next section the process window of the basic process strategy is extended and finally the modified process strategies are studied.
Finally, the numerical simulation results are summarized and future research topics outlined.
\section{Numerical Methods}
%
%
Thermodynamic fluid transport including phase transitions is modeled by single phase-continuum conservation equations.
These are the incompressible Navier-Stokes equations including a force term for the gravity and the advection diffusion equation for the energy density including a a source term for the absorption of the electron beam.
This simplified model neglects fluid compression, viscous heat dissipation and all evaporation effects, like material and energy loss and recoil pressures.

The Navier-Stokes equations are solved by an isothermal LB method introduced by~\citeauthor{McNamara1988}~\cite{McNamara1988} and \citeauthor{Higuera1989}~\cite{Higuera1989}.
The idea of a LB method is solving the Boltzmann equation in the hydrodynamic limit for a particle distribution function (pdf) in the physical momentum space. 
A pdf describes the probability of finding a particle with a microscopic velocity at a certain position and time.
The local macroscopic values of the density and the velocity are evaluated by integration.
In the incompressible flow limit, i.e. for small Mach numbers, the single phase-continuum conservation equations can be derived by a Chapman-Enskog expansion~\cite{Palmer2000,Chatterjee2006,Shi2009}.

LB methods, which additionally solve the advection diffusion equation, are the multi-speed and multi-distribution approach. 
The multi-distribution LB method avoids the multi-speed drawbacks of unstable simulations limited to one Prandtl number using a separate pdf for the energy density.
\citeauthor{Massaioli1993}~\cite{Massaioli1993} and \citeauthor{Shan1997}~\cite{Shan1997} successfully use this approach to simulate the Rayleigh-Bernard convection at moderate and near-critical Rayleigh numbers.
\citeauthor{He1998}~\cite{He1998} extend this approach, validate it by the Couette flow and the Rayleigh-Bernard convection
and conclude that this approach describe the viscous heat dissipation and compression work correctly and indicates better stability in contrast to multi-speed approaches.

The electron beam as the energy source for the advection diffusion equation is modeled by a two-dimensional Gaussian distribution~\cite{Markl2013}.
The standard deviation as well as the full width half max (FWHM) diameter, the width of the Gaussian distribution at half of the maximum beam power, are defined in the simulation setup.
The absorption within the material is realized by an exponential Lambert-Beer law described by \citeauthor{Markl2013}~\cite{Markl2013}

\citeauthor{Koerner2005}~\cite{Koerner2005} introduce a volume-of-fluid based free surface lattice Boltzmann approach, where the gas phase and the material are separated by a closed layer of interface cells.
Each interface cell is partly covered by material, defined by a fill level. 
The gas phase is neglected, thus the method is only applied on material cells assuming that the thermodynamic behavior is completely covered. 
Boundary conditions imposed at the interface ensure mass, momentum and energy conservation~\cite{Ammer2013}.

The powder bed generation algorithm~\cite{Ammer2013} is executed at the beginning of each simulation.
The initialized powder particles in a regular grid execute a free fall onto the previous layer until they reach a steady state.
Subsequently, the particles are coupled into the LB approach by modifying the fill levels and initializing missing macroscopic values and pdfs.
The powder particle size distribution is approximated by an inverse Gaussian distribution~\cite{Ammer2013}, 
between a minimum and maximum diameter using a skewness parameter, describing the asymmetry of the distribution function from the mean value.
%
%
\section{Simulation Results}
%
%
In this section we describe an experimental setup which is used to define the numerical simulation setup.
The limitations of the basic process strategy are determined by the extension of the numerical process window by numerical simulations up to scan velocities of 30\,m/s.
Subsequently, two modified process strategies are examined: first, we increase the beam diameter and second, we decrease the line offset, whereby the scan velocity is always increased in order to reduce the total build time.
%
%
\subsection{Experimental and numerical setup}
%
%
In the experimental setup~\cite{Juechter2014252} nine cuboids of $15\,\,\text{mm}\times15\,\text{mm}\times10\,\text{mm}$ are generated by hatching Ti-6Al-4V powder particles on a start plate with a line offset of 0.1\,mm.
The build platform is lowered for each new layer about 0.05\,mm resulting in an effective layer thickness between 0.08\,mm and 0.10\,mm, due to the densification during melting.
The scanning direction from layer to layer is rotated by 90$^\circ$.
In all experiments the preheating temperature is 650$^\circ$C and gas atomized Ti-6Al-4V powder with a size distribution ranging from 0.045\,mm to 0.105\,mm and a mean value of 0.069\,mm is used.
The hatching differs by line energy and scan velocity of the electron beam. 
The line energy $E_{L}=UI/\lVert\underline{u}_\text{b}\rVert$, where $U$ denotes the acceleration voltage, $I$ the beam current and $\lVert\underline{u}_\text{b}\rVert$ the absolute scan velocity of the electron beam. 
The parameter set $(E_{L},\lVert\underline{u}_\text{b}\rVert)$ defines the configuration, whereby the acceleration voltage is fixed to 60\,kV. 
%
\begin{figure}[btp]
\centering
\includegraphics[width=\textwidth]{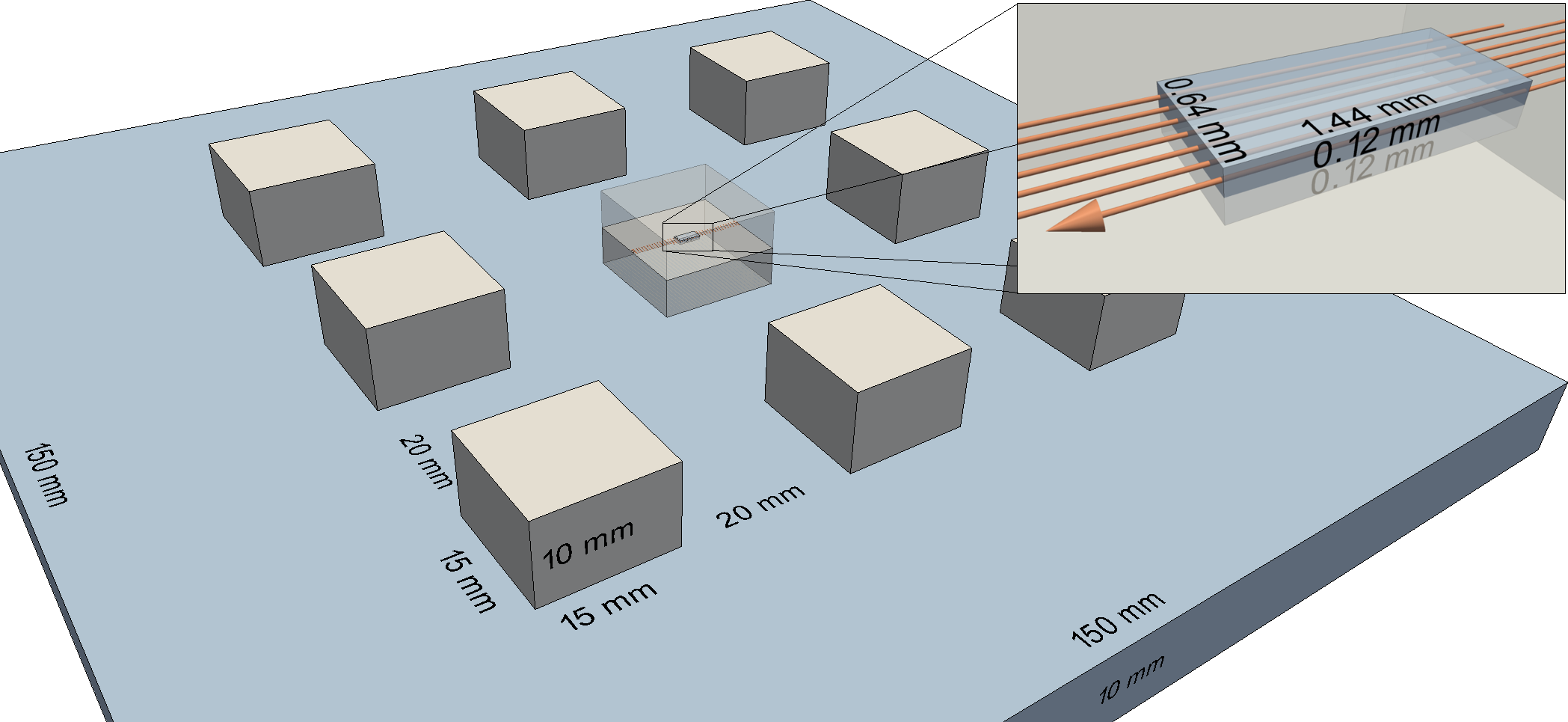}
\caption{Experimental Setup: Nine cuboids of $15\,\,\text{mm}\times15\,\text{mm}\times10\,\text{mm}$ are hatched with a line offset of 0.1\,mm and layer thickness of 0.05\,mm; 
         Simulation Setup (top right): A cuboid of $1.44\,\text{mm}\times0.64\,\text{mm}\times0.24\,\text{mm}$ located in the center of a cuboid is simulated with seven scan lines;}
\label{fig:setup}
\end{figure}

Because of the high computational costs of three dimensional simulations, we only model the hatching of one powder layer with an effective layer thickness of 0.1\,mm instead of multiple layers. 
We also minimize the simulated powder particle layer, focusing on a domain of $1.44\,\text{mm}\times0.64\,\text{mm}\times0.24\,\text{mm}$ (cf. \autoref{fig:setup} (top right)).
The previous layer is approximated by completely dense material with a height of 0.12\,mm.
Seven scan lines with an offset of 0.02\,mm from the domain boundary and a line offset of 0.1\,mm are simulated. 
The corresponding numerical powder size distribution, defined in the same value range than the real size distribution, uses a mean value of 0.061\,mm and a skewness parameter of 0.809.
The electron beam standard deviation is 0.1\,mm, which corresponds to a FWHM diameter of 0.235\,mm.
The lattice spacing is $\Delta x=5\cdot10^{-6}$\,m and the time step is $\Delta t = 1.75\cdot10^{-7}$\,s.
At the bottom of the simulation domain, a thermal Dirichlet boundary condition~\cite{Ginzburg20051196} with a build chamber temperature of 650$^\circ$C 
and a no-slip half-way bounce-back boundary condition are used.
In both other dimensions periodic boundary conditions are applied, to influence the melt pool dynamics as less as possible.
In order to minimize numerical errors induced by the boundary treatment, all measurements and images cut off the outside regions by 0.1\,mm.
All material parameters are directly taken from the properties of Ti-6Al-4V~\cite{Boivineau2006507, boyer1998, asm1990metals, Lu2005}.
\citeauthor{Ammer2014}~\cite{Ammer2014} justifies these simplifications of the numerical setup by the good agreement between the experimental and numerical process windows up to 6.4\,m/s.

The numerical results are classified into `porous', `good' and `uneven' in accordance to the experimental classification~\cite{Juechter2014252}.
Due to the volume of fluid approach, the relative density of the numerical domain is computed by the ratio of the material volume to the total subsurface volume.
The upper bound for `porous' results is a relative density of 99.5\%.
Because the numerical simulation excludes evaporation effects, no significant top surface distortions occur, due to missing evaporation recoil pressures.
Therefore, a numerical simulation is classified as `uneven', when the averaged maximum melt pool temperatures exceed 7500\,K~\cite{Ammer2014}.
Using this threshold, we assume, that the top surface geometry is directly related to this temperature.
Because of the missing cooling effect caused by evaporation, this temperature limit is not comparable with real experiments, where maximum temperatures lower than 4000\,K are observed~\cite{Klassen201402}.
All other simulation results between these bounds are declared as `good'.
%
%
\subsection{Process Window Extensions}
All results in the course of this paper are solely numerical results, due to the beam power restrictions of the available SEBM machines.
Constant and reliable beam characteristics are only available up to 1.2\,kW~\cite{Juechter2014252}.
Considering an exemplary line energy of 0.1\,kJ/m, this threshold is already reached at a scan velocity of 12\,m/s.
In order to decrease build times, the scan velocity is generally increased causing higher beam powers, which are currently not machinable.
All numerical configurations are generally simulated once, except some configurations near the upper and lower bound, in order to verify the correct classification.

In this section we extend an experimentally validated numerical process window ranging up to scan velocities of 6.4\,m/s~\cite{Ammer2014}.
While increasing the scan velocity we study the trend of the porosity and evaporation bounds to determine the fastest configuration of scan velocity and line energy.

\begin{figure}[btp]
\centering
 \subfigure[$t=0.24$\,ms]{\includegraphics[width = 0.46\textwidth]{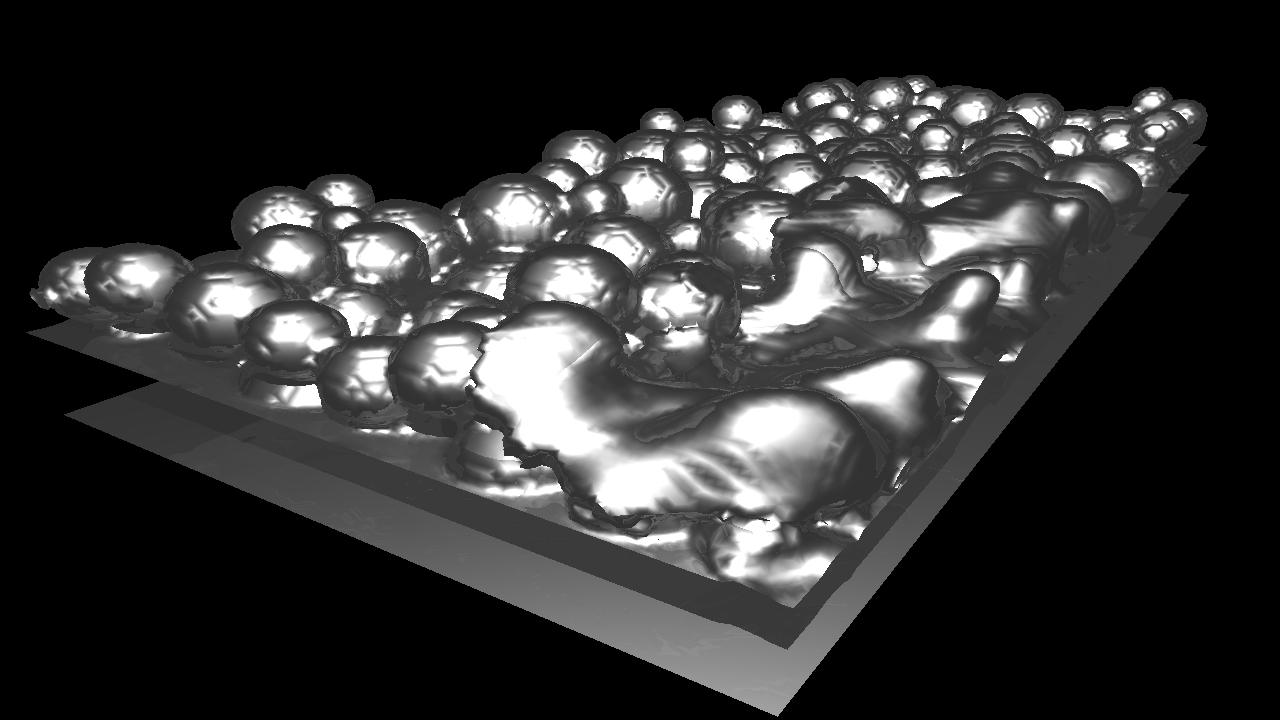}\label{fig:Hatching:a}}  \hfill
 \subfigure[$t=1.24$\,ms]{\includegraphics[width = 0.46\textwidth]{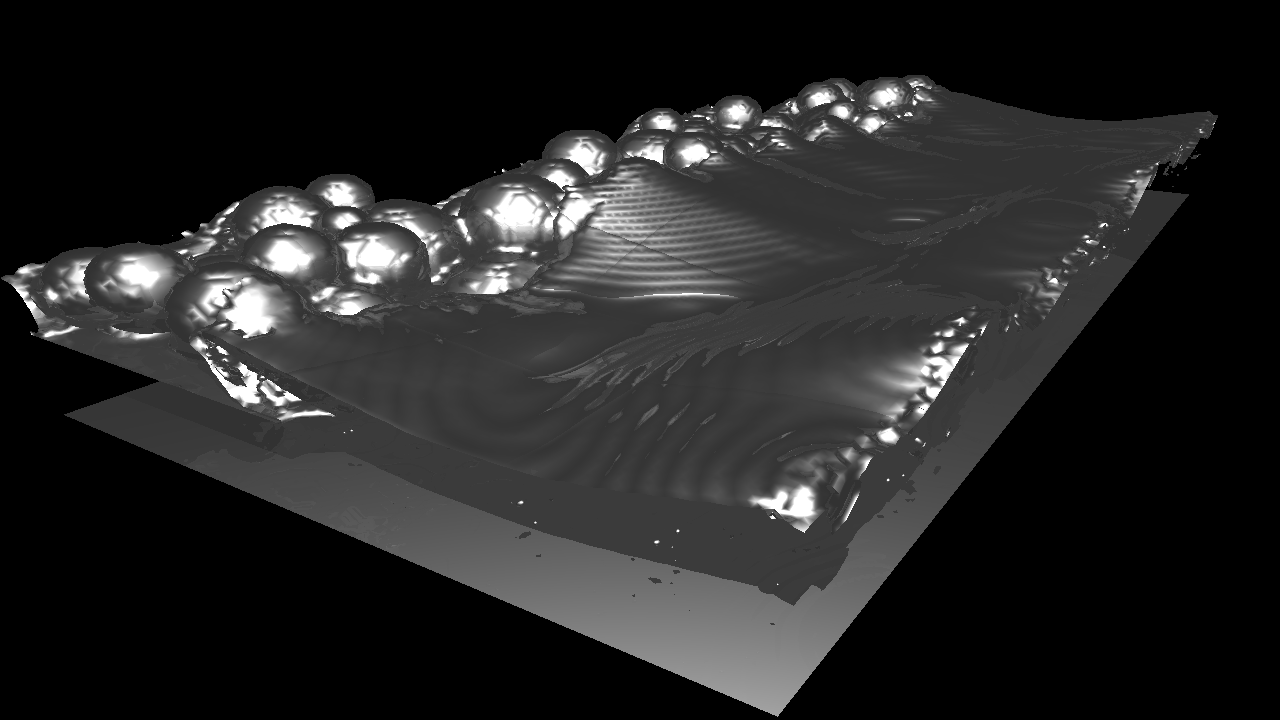}\label{fig:Hatching:b}}  \hfill
 \subfigure[$t=2.24$\,ms]{\includegraphics[width = 0.46\textwidth]{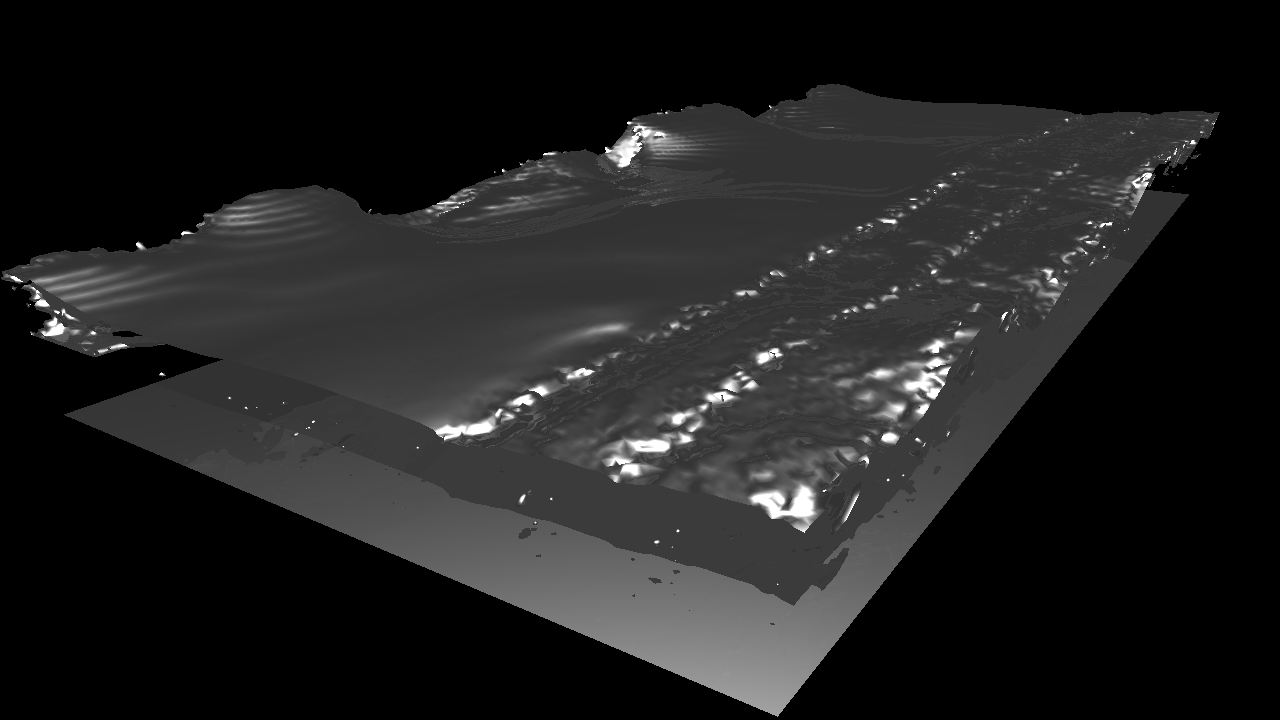}\label{fig:Hatching:c}}  \hfill
 \subfigure[$t=3.24$\,ms]{\includegraphics[width = 0.46\textwidth]{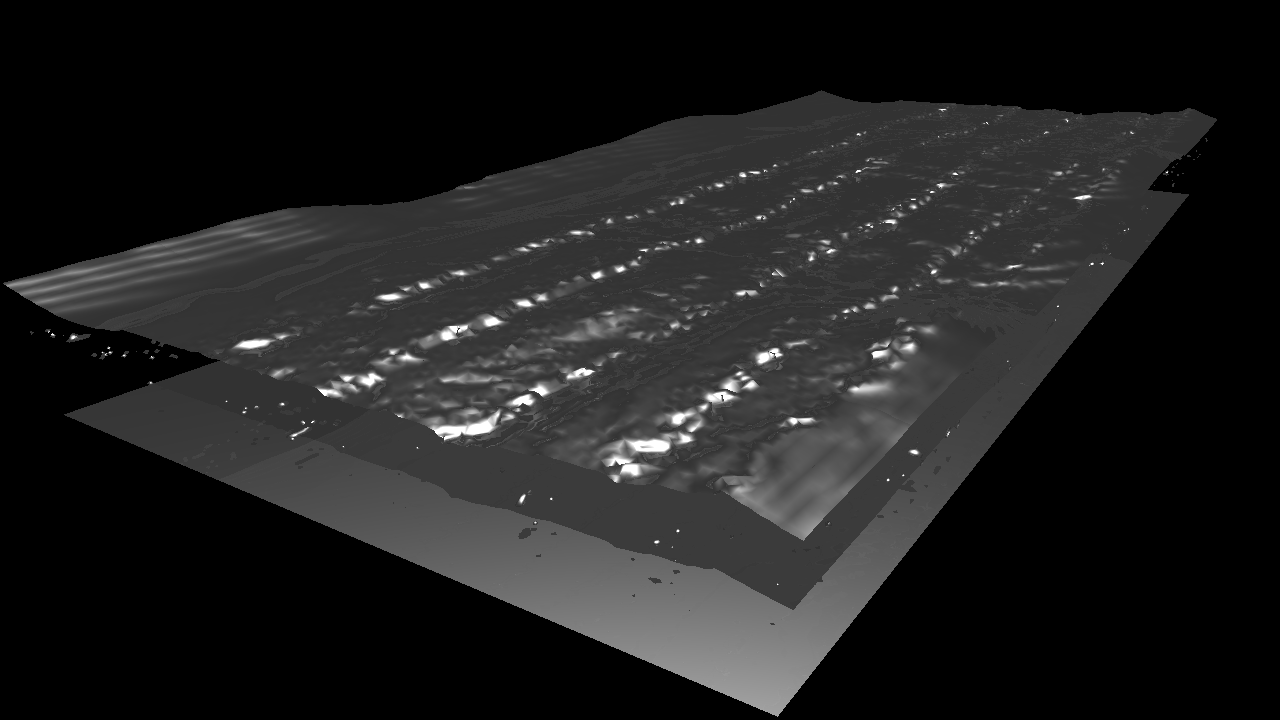}\label{fig:Hatching:d}} 
\caption{Ray traced images of hatching one layer with a line energy of 0.10\,kJ/m and a scan velocity 15\,m/s; 
fill level isosurfaces of the top surface and the bottom plate are shown;
liquid melt pool has a very smooth surface; solidified material show grooves and more scattering surface;}
\label{fig:Hatching}
\end{figure}

\autoref{fig:Hatching} shows ray traced images\footnote{Ray traced images generated with povray: \url{http://www.povray.org/}} of hatching one layer with the configuration set (0.1\,kJ/m, 15\,m/s) at four different time steps. 
The free surface of the powder particles is visualized by an isosurface on the fill level.
In~\autoref{fig:Hatching:a} the electron beam is located in the center of the first scan line and the affected powder particles are not yet completely melted. 
\autoref{fig:Hatching:b} shows the smooth melt pool after the third scan line is processed. 
On the left unaffected particles, which are partly covered by the melt pool, and a groove at the right end of the melt pool are visible.
In~\autoref{fig:Hatching:d} the electron beam scanned the whole domain, which is already resolidified, indicated by the grooves on the free surface. 
They grow each time the melt pool starts solidifying from the outer regions due to wetting effects.
There are no obvious unmelted particles or layer bonding defects between the bottom plane and the original powder layer.
However, the relative density is less than 99.5\% which cause a classification to `porous'.

\begin{figure}[btp]
\centering
\includegraphics[scale=0.375]{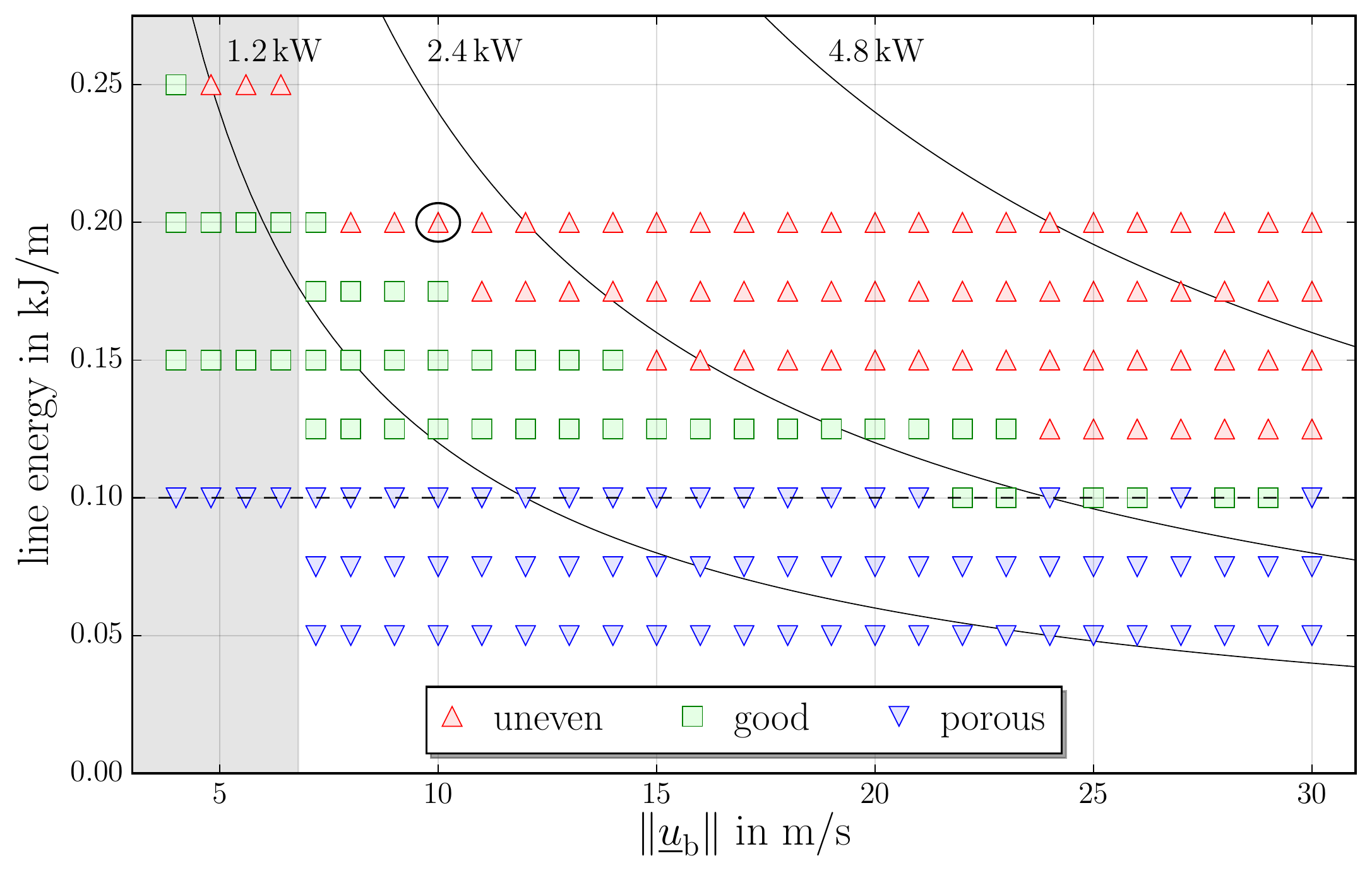}
\caption{Extended numerical process window with 0.1\,mm line offset: porosity bound nearly constant, while evaporation bound decreases, which results in a process window closing at least at 30\,m/s; 
         gray area indicate experimentally validated numerical results; dashed line at 0.1\,kJ/m and circled configuration (0.2\,kJ/m, 10\,m/s) are compared in the following sections;}
\label{fig:pw}
\end{figure}

\autoref{fig:pw} shows the classifications of samples with different line energies and scan velocities of the electron beam, including the previous example.
Configuration sets causing `porous', `good' and `uneven' results are represented by blue downward oriented triangles, green squares and red upward orientated triangles, respectively.
The gray area indicate results of scan velocities up to 6.4\,m/s, which has been compared to experimental data~\cite{Ammer2014}.

The last numerical simulation classified as `good' uses a scan velocity of 29\,m/s, thus the process window is finally closed at 30\,m/s.
The reason is the nearly constant lower porosity bound, while the upper evaporation bound decreases.
Higher scan velocities above 30\,m/s cause configurations, where the results are `porous' as well as `uneven', 
because the energy input is insufficient to melt the complete powder layer but the locally absorbed beam energy causes high top surface distortions.

The upper bound is defined by the maximum top surface temperature.
Increased scan velocities with a constant line energy cause a deposition of the same total energy in a shorter duration.
As a first consequence, the resting temperature in the melt pool is higher when the electron beam returns in the next scan line.
This effect causes an increasing maximum temperature at the top surface in the center of the melt pool from scan line to scan line.
Thus, the linear increasing deposited energy with the scan velocity overlaps this hatching effect.
The maximum temperature is well predictable, which cause a reliable upper bound to `uneven' results.

The lower bound stays nearly constant, because the total amount of energy to melt the layer only depends on the volume of the powder layer.
However, all seven scan lines are performed on the order of milliseconds, where only a small fraction of the deposited energy is lost by diffusion into the base plate.
Due to less energy loss with higher scan velocities, the lower bound is slightly decreasing.
Besides this trend, the lower bound show alternating classifications between `porous' and `good', especially for a line energy of 0.1\,kJ/m (\autoref{fig:pw}, dashed line).
Even if the minimum total energy to melt the whole layer is deposited into the material, the appearance of the powder bed has a statistical fluctuation.
Thus, there exist powder bed compositions, where the heat transfer is hampered.
This is a critical fact for a process strategy, because it should lead to reproducible results with low deviations.
Concluding, a reliable configuration set needs an offset from the lower bound to ensure that statistical fluctuations in the powder bed do not affect the final result.

Therefore, a practical process window will close at latest at 20\,m/s with a line energy of 0.125\,kJ/m.
Comparing the maximum applied beam power of 2.5\,kW with the possibilities of a 10\,kW electron beam gun, there is a huge potential to improve this process strategy.
%
%
\subsection{Increasing Beam Diameter}
A possible strategy to use a higher beam power is to increase the beam diameter. 
The peak power density at the center of the electron beam decreases, because the energy distribution of the electron beam is modeled by a two dimensional Gaussian distribution, resulting in lower maximum temperatures at the top surface.
With the reduced maximum temperatures, we are able to increase the line energy and accelerate the build velocity by using higher beam powers.

We study two additional beam shapes with increased beam diameters, where the affected area is increased by 50\% and 100\%.
The resulting standard deviations are 0.122\,mm and 0.141\,mm with a FWHM of 0.287\,mm and 0.332\,mm, respectively.

\begin{figure}[btp]
\centering
\subfigure[Increased affected area of  50\%]{\label{fig:pw-b0}\includegraphics[scale=0.375]{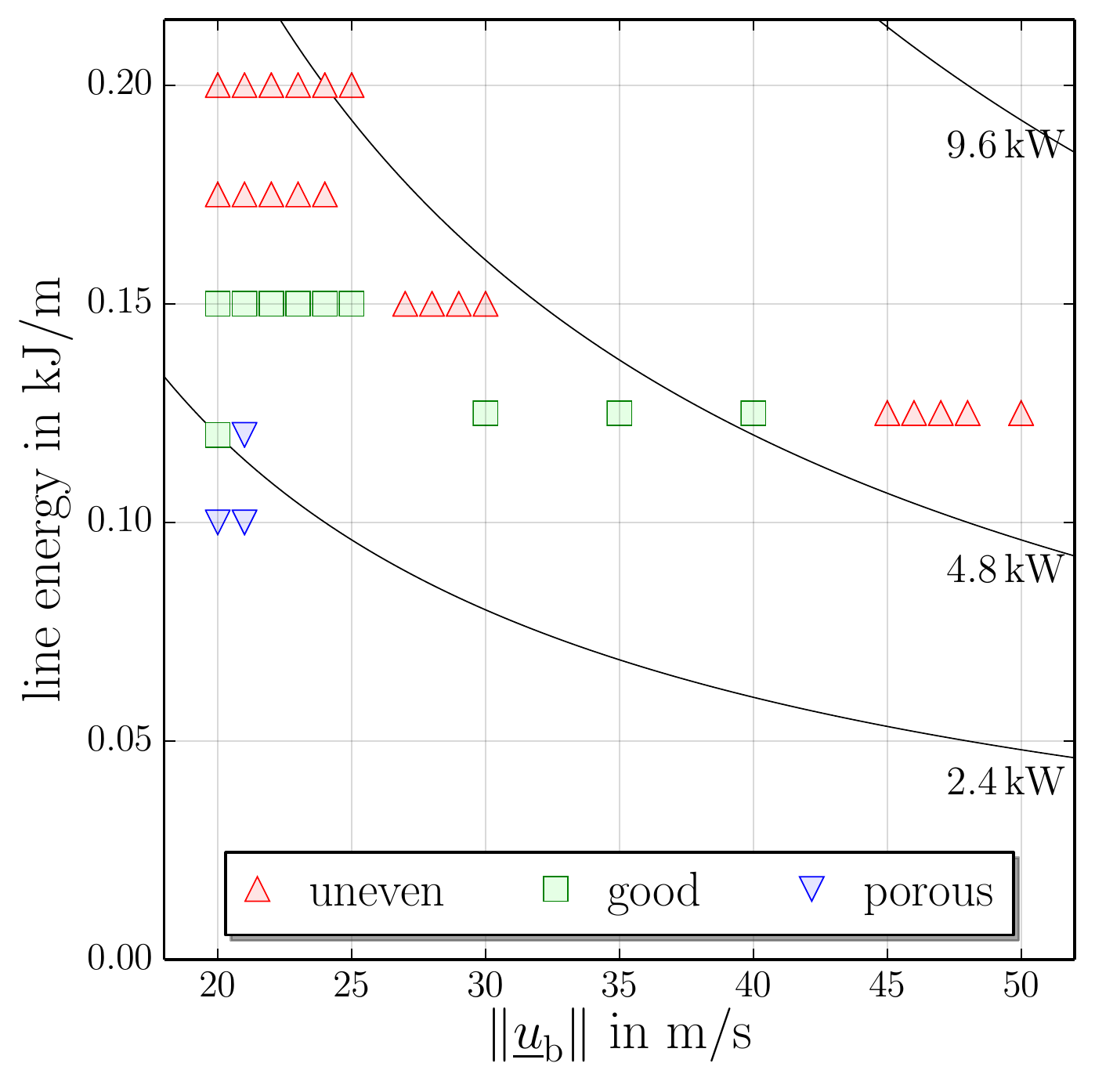}} \hfill
\subfigure[Increased affected area of 100\%]{\label{fig:pw-b1}\includegraphics[scale=0.375]{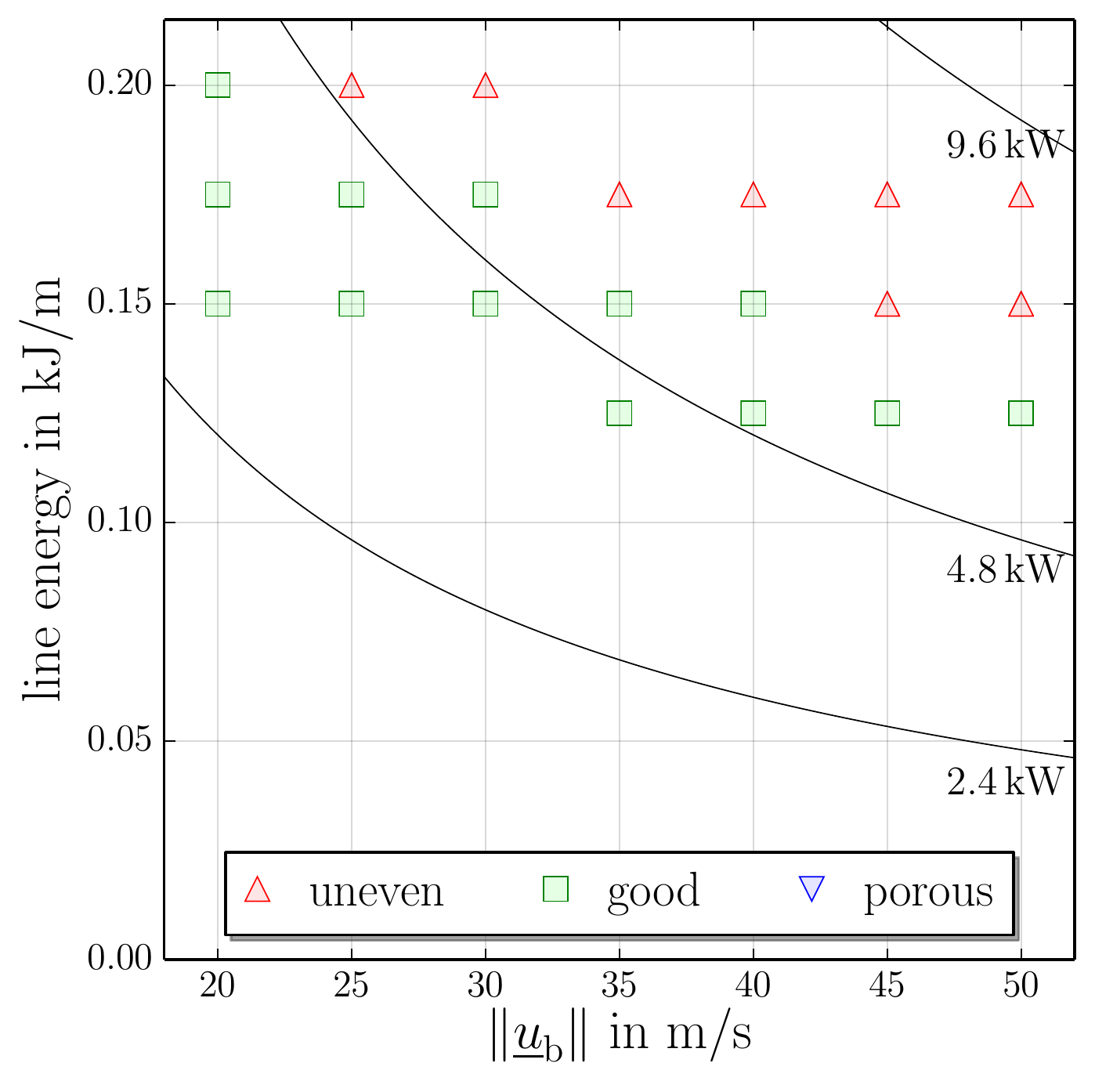}} 
\caption{Partial process windows with increased beam diameters: usable beam power of at least 5\,kW (a) and 6.25\,kW (b) with a possible build time reduction of 60\% and more}
\label{fig:pw-b}
\end{figure}

\autoref{fig:pw-b} shows simulation results for both beam diameters.
When the affected area is increased by 50\%, then a line energy of 0.125\,kW with a scan velocity of 40\,m/s is possible, which result in a total beam power usage of 5\,kW (cf. \autoref{fig:pw-b0}).
Enlarging the affected are by 100\% result in a maximum configuration of (0.125\,kW, 50\,m/s) with a total beam power usage of 6.25\,kW (cf. \autoref{fig:pw-b1}).
Thus, total build time reductions with a scan velocity of 50\,m/s, compared to 20\,m/s with the basic strategy, of 60\% are achievable.

These results justify that the potential of 10\,kW electron beam guns can be applied with this strategy.
Nevertheless, this strategy can cause new challenges, especially at the horizontal borders of the geometry.
The resulting beam diameter leads to an increased melt pool width, which can deteriorate the dimensional accuracy.
To circumvent this drawback, it is possible to melt the contour separately, before or after hatching.
The hatching time for one complete layer of a cuboid with the basic process strategy at 20\,m/s is 112.5\,ms.
Suppose an increased beam diameter and using a scan velocity of 40\,m/s this duration is reduced to 56.25\,ms.
However, contouring the hatch with a slow electron beam of 0.5\,m/s results in an additional time of 120\,ms.
Thus, the additional contour scanning time destroys the build time reduction, because the contour has to be melted with a slow scan velocity.
A possible technique to reduce contouring times and restore the total build time reduction is the multi-beam approach, where different locations are melted simultaneously.

Summarizing, this strategy is only useful when the hatching regions have a simple geometry, large area and when no additional contour scanning is necessary.
In a post-processing step the dimensional accuracy can be further improved.
%
%
\subsection{Decreasing Line Offset}
Another strategy to decrease build time and use higher beam powers is to decrease the line offset of the scan lines. 
Although the total scan length increases, the scan velocities are adjustable by a higher factor, which reduces the total build time.
In this case, the line offset as well as the line energy are halved to keep the total electron beam energy constant. 
This results in thirteen scan lines within the numerical domain with the same initial offset of 0.02\,mm.
The configuration sets are comparable to the basic process strategy, when the scan velocity is doubled.

\begin{figure}[btp]
\centering
\includegraphics[scale=0.375]{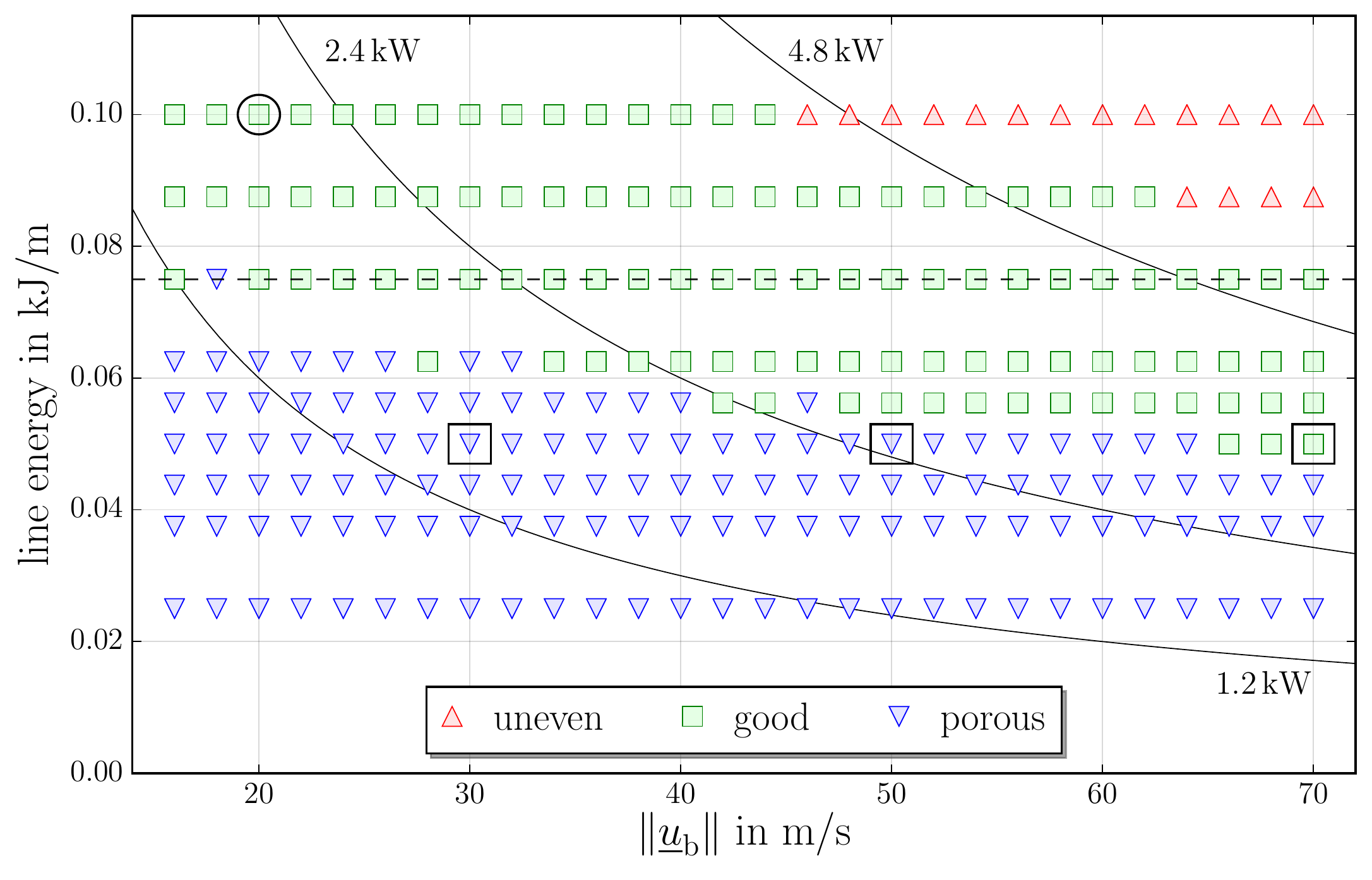}
\caption{Numerical process window with a scan line offset of 0.05\,mm: usable beam powers of at least 5\,kW with build time reductions of at least 43\%, e.g., with a configuration set of (0.075\,kJ/m, 70\,m/s);
         circled configuration (0.1\,kJ/m, 20\,m/s) comparable to equally marked configuration in \autoref{fig:pw};
				 dashed line and squared configurations are compared in the following sections;}
\label{fig:pw-lo}
\end{figure}

\autoref{fig:pw-lo} shows the numerical results for this strategy up to 70\,m/s. 
The upper and lower bounds show the same trends as in \autoref{fig:pw}.
With this strategy the usable beam powers are easily shifted beyond 5\,kW.

The basic principle, why higher beam powers are usable, is the same than with the increased beam diameters.
Due to the faster scan velocity, the energy of the electron beam is deposited on a larger area compared to the basic strategy within the same time.
Thus, the mean temperature of the beam affected material stays on a high level over the complete scan length, whereby the maximum temperatures decrease.
This result in a fundamental change of the melt pool shape from a local drop shape to a shape, where the material along the complete scan line becomes liquid.
Due to these facts, the melt pool is less dynamic and have lower maximum evaporation rates.

The link to \autoref{fig:pw}, indicated by the circled configurations, is found by halving the scan velocity and doubling the line energy, to keep the used beam power and build time constant.
The current configuration of (0.1\,kJ/m, 20\,m/s) result in a fully dense part with a smooth top surface, whereas the corresponding configuration with the basic process strategy at (0.2\,kJ/m, 10\,m/s) cause an uneven top surface.
Thus, with the halved line offset strategy, it is possible to increase the scan velocities by a higher factor than two in order to decrease the total build time.

\begin{figure}[btp]
\centering
\subfigure[0.10\,mm line offset]{\label{fig:mt-100}\includegraphics[scale=0.375]{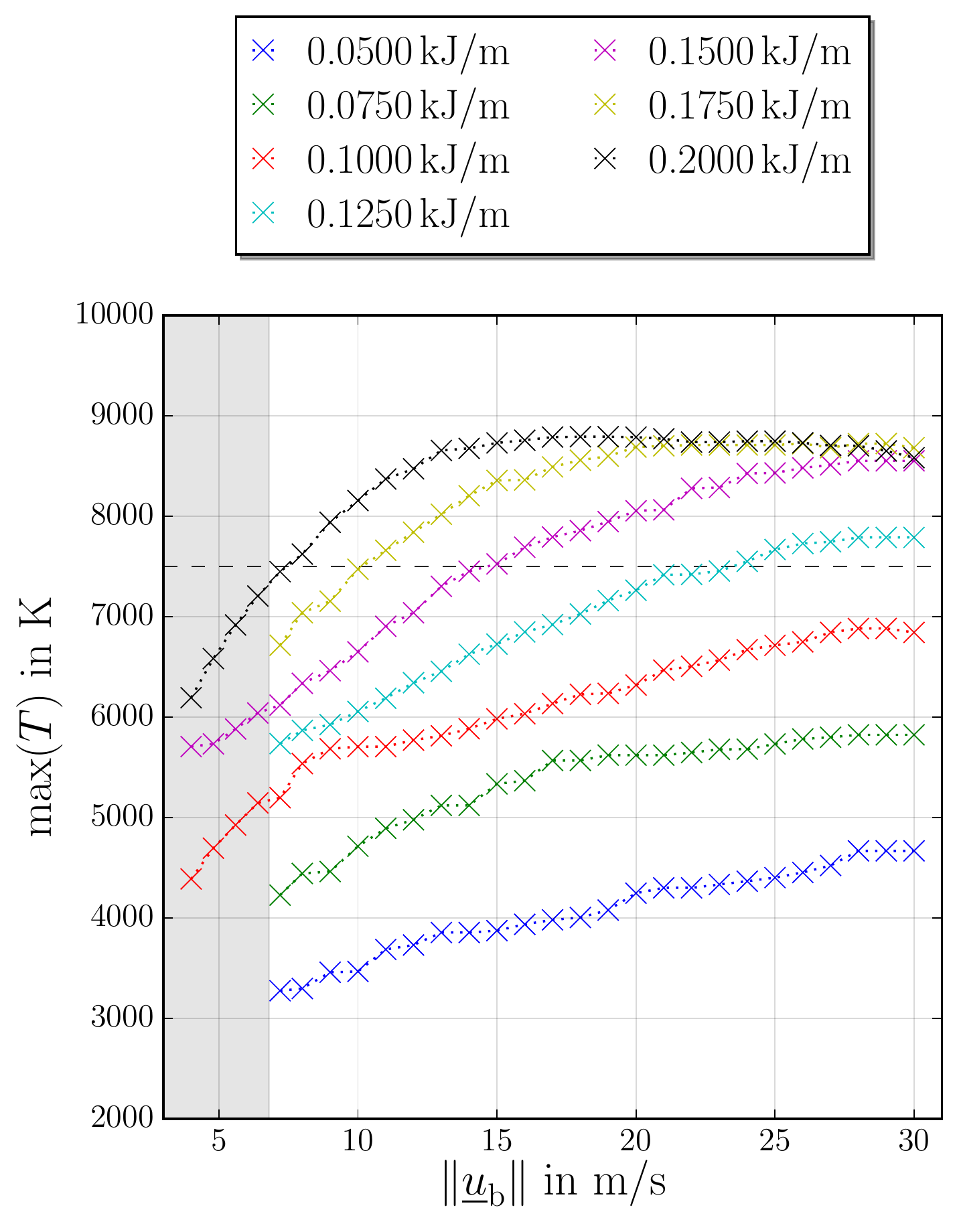}} \hfill
\subfigure[0.05\,mm line offset]{\label{fig:mt-050}\includegraphics[scale=0.375]{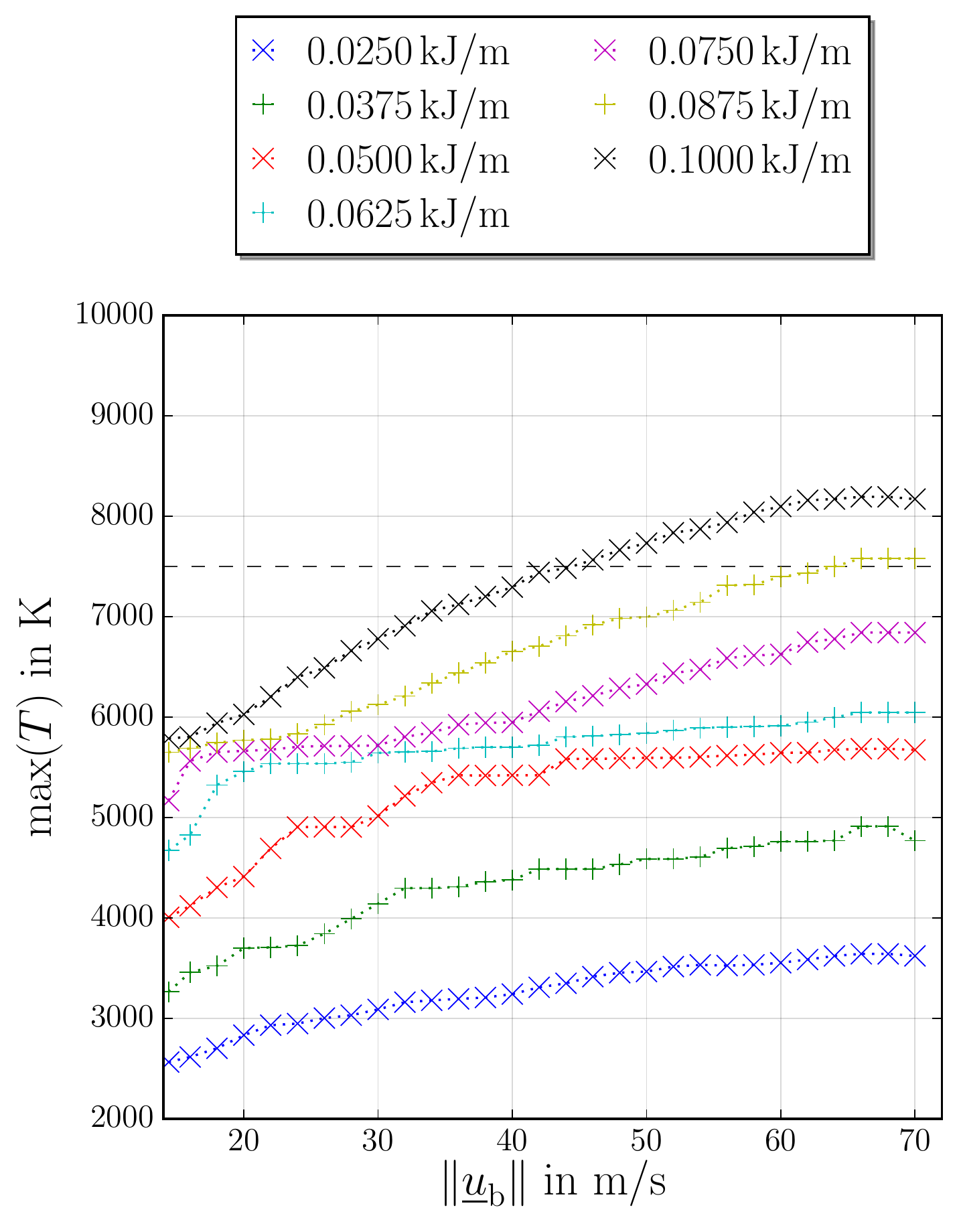}} 
\caption{Maximum temperatures for basic (a) and decreased line offset (b) strategies: 
         colored lines are pairwise comparable with the same beam power using doubled speed for halved line offset;
				 gray area indicate experimentally validated numerical results (a);
         decreased line offset strategy significantly decrease maximum temperatures for comparable setups, e.g., 
         results for 0.075\,kJ/m (b) are generally under the temperature threshold, whereby for 0.150\,kJ/m (a) the temperature threshold is reached at 15\,m/s;}
\label{fig:mt}
\end{figure}

The maximum temperatures of the decreased line offset and the basic process strategy are compared in \autoref{fig:mt}.
The colored lines are pairwise connected by the same beam power using the doubled speed and the halved line offset.
A maximum line energy of approximately 0.1\,kJ/m is applicable to reach smooth top surface parts with high scan velocities for the basic process strategy (cf. \autoref{fig:mt-100}).
However, compared with \autoref{fig:pw} (dashed line), almost all parts are classified as `porous'.

Comparing the corresponding line energy of 0.05\,kJ/m in \autoref{fig:mt-050}, the temperature threshold is not reached.
It is even possible to increase the line energy up to 0.075\,kJ/m without generating `uneven' results.
In comparison with \autoref{fig:pw-lo} (dashed line), nearly all simulations result in fully dense parts.

\begin{figure}[btp]
\centering
\includegraphics[scale=0.375]{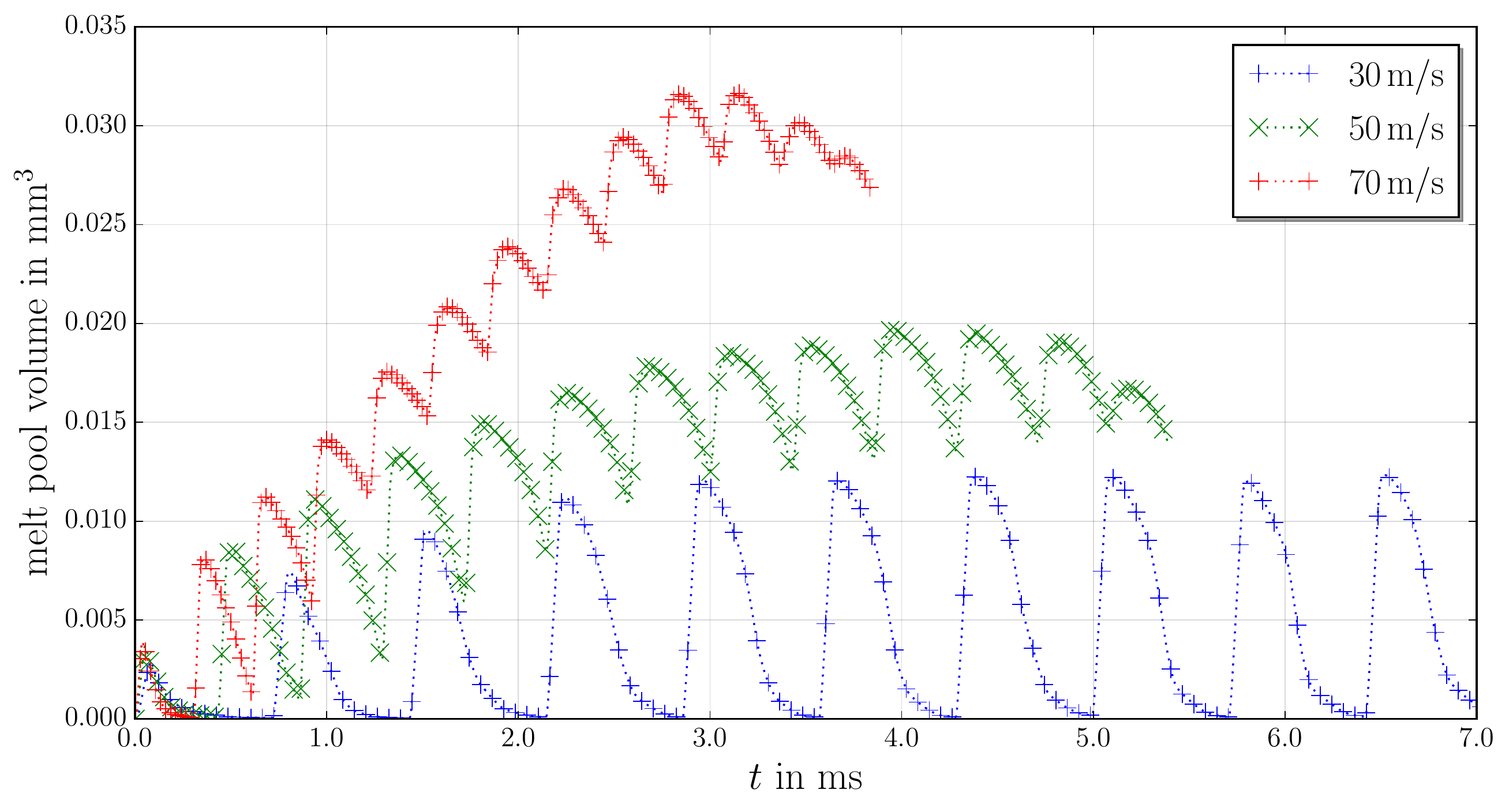}
\caption{Time dependent melt pool volume for different scan velocities with 0.05\,kJ/m: 
				 At a scan velocity of 50\,m/s and 70\,m/s the previous scan line is still liquid and the melt pool grows continuously}
\label{fig:vd}
\end{figure}

The temporal evolution of the melt pool lifetime is shown in \autoref{fig:vd}, 
where the melt pool volume depending on the simulation time for different scan velocities with 0.05\,kJ/m is shown.
This line energy is chosen, because nearly all simulations result in `porous' parts, except the fastest configurations cause fully dense parts.
The link of these configurations to \autoref{fig:pw-lo} is established by the rectangle marked symbols.
The melt pool size increases with increasing scan velocity, whereas the lifetime stays nearly constant. 
In the first example at 30\,m/s, the melt pool is already completely solidified when the electron beam returns.
Thus, there is no benefit by the hatching technique and the result is a porous part.
Increasing the scan velocity to 50\,m/s, the melt pool is available over the whole simulation duration.
This indicates a change in the melt pool geometry, from a drop shaped form with a much shorter length than the scan line length and a large width towards a rectangular geometry with a small width, where the complete scan line length is liquid.
It reaches a maximum volume after ten scan lines, which is 1.6 times larger than the previous example.
This maximum is reached, because the next scan lines are only partially inside the domain and melt less particles.
Nevertheless, the trend in the melt pool growth indicate, that the natural maximum volume with this scan velocity is almost reached.
Furthermore, the melt pool depth is not sufficient to achieve a fully dense part by remelting the top fraction of the previous layer.
The last scan velocity at 70\,m/s, show the same tendency, but reach a 2.6 times larger total melt pool volume, which finally results in a dense part.
After the first three scan lines, the growth per scan line of the melt pool is nearly constant,
which indicate, that the maximum melt pool volume is only reached due to the domain border.

Summarizing, this strategy increases the width of the processing window significantly
and results in more reliable and reproducible processes that have a reduced build time of at least 43\% by making use of the high energy potential of the electron beam.
The dimensional accuracy is expected to stay the same compared to the basic process strategy, because the beam diameter remains the same.
%
%
\section{Conclusions}
We have been presented a numerical method to simulate the SEBM process and have been applied it on hatching process strategies of cuboids.
A numerical process window for a state-of-the-art process strategy has been extended to higher scan velocities.
It has been shown that this process window closes at approximately 20\,m/s with a total beam power usage of 2.5\,kW,
because a safety margin from the lower bound of the process window is necessary to manufacture multiple parts with equal quality.

In order to use the high potential of future electron guns with a total beam power of 10\,kW, we have been studied different hatching strategies.
The first technique with increased beam diameters results in a process window, where higher beam powers of at least 5\,kW are applicable with reduced build times of at least 60\%.
However, this strategy encloses a potential risk that the dimensional accuracy deteriorates, due to a wider melt pool.
An additional contouring step before or after hatching is able to improve the dimensional accuracy but lowers the build time advantage.
The second hatching strategy using a decreased line offset and increased scan velocities also result in a deposition of the electron beam energy on a larger area.
This effect cause similar process improvements with a beam power usage beyond 5\,kW at 70\,m/s scan velocity with a build time reduction of at least 43\%, 
whereby the dimensional accuracy is not lowered, because the beam diameter is equal to the basic process strategy.

With both process strategies we conclude, that modified scan strategies are able to use the potential of future electron beam gun powers 
to decrease the build time, reduce manufacturing costs and extend the variety of possible applications and parts.
Certainly, there are many other parameters which are adjustable to achieve similar process strategy improvements including the powder particle size distribution, layer thickness or acceleration voltage.
However, besides further improvements of enhanced process strategies for hatching, the inclusion of an evaporation model~\cite{Klassen201402} is necessary to improve the prediction of `uneven' top surfaces.
The results of this paper are not directly applicable to the manufacturing process without an experimental validation, but they highlight possible advantageous process strategy modifications.

%
%
\begin{acknowledgements}
Our work is supported by the European Union Seventh Framework Program -- Research for SME's with full title 
``High Productivity selective electron beam melting Additive Manufacturing Development for the Part Production Systems Market'' and grant agreement number 286695.
\end{acknowledgements}
%
%
\bibliographystyle{spbasic}      
\bibliography{literature}   
\end{document}